\documentclass[aps,graphicx,twocolumn,epsfig]{revtex4}

\usepackage{epsfig}

 \draft

\begin{document}

\title{Propagation of sound in a Bose Einstein condensate in an
optical lattice} 

\author{C.~Menotti$^1$, M.~Kr\"amer$^1$, A.~Smerzi$^{1,2}$, 
L.~Pitaevskii$^{1,3}$, and S.~Stringari$^1$} 
\affiliation{
$^1$Istituto Nazionale per la Fisica della Materia BEC-CRS 
and Dipartimento di Fisica, Universit\`{a} di Trento, 
I-38050 Povo, Italy\\
$^2$Theoretical Division, Los Alamos National Laboratory, 
Los Alamos, NM 87545, USA \\
$^3$ Kapitza Institute for Physical Problems, 117334 Moscow, Russia}

\date{\today}

\begin{abstract}
We study the propagation of sound waves in a Bose-Einstein condensate
trapped in a one-dimensional optical lattice.  We find that the
velocity of propagation of sound wavepackets decreases with increasing
optical lattice depth, as predicted by the Bogoliubov theory.  The
strong interplay between nonlinearities and the periodicity of the
external potential raise new phenomena which are not present in the
uniform case. Shock waves, for instance, can propagate slower than
sound waves, due to the negative curvature of the dispersion
relation. Moreover, nonlinear corrections to the Bogoliubov theory
appear to be important even with very small density perturbations,
inducing a saturation on the amplitude of the sound signal.
\end{abstract}

\maketitle

The study of Bose Einstein condensates in optical lattices is a very
active field of research, both from the theoretical and experimental
sides.  The presence of the lattice can drammatically modify the
behaviour of the system with respect to the uniform case, giving rise
to new phenomena like, for instance, a Mott-insulator superfluid phase
transition \cite{jaksch, greiner} or the occurence of dynamical
instabilities \cite{wu2001a, smerzi2002, menotti2003, cataliotti2003,
modugno2003a, cristiani2003a}.  
Several efforts are also 
focusing on the creation of
quantum computers \cite{rolston02} and ultrasensitive interferometers
\cite{kasevich02}.
In this paper we study the propagation
of sound waves on top of a Bose-Einstein condensate at rest in a
one-dimensional lattice.

The propagation of sound in a harmonically trapped condensate without
lattice has been already observed experimentally
\cite{andrews1997a,andrews1998a} and studied theoretically
\cite{zaremba1997a,kavoulakis1997a,stringari1998a,damski2003a}. Generally
speaking, it is important to study the propagation of sound also in
the non linear regime, where density fluctuations are comparable to
the background density of the condensate. The reason is that only
large amplitude wavepackets can be realistically observed
experimentally.  It turns out, however, that nonlinear effects in the
sound propagation are particolarly interesting also from the
theorethical point of view.  For instance, the formation of shock
waves in front of a bright sound wavepacket (positive density
variation) a uniform or harmonically trapped condensates has been
studied in \cite{damski2003a}, and analitycal solutions have also been
discussed in \cite{gurevich1973a,gurevich1987a,note}. In the presence
of a periodic potential, theoretical aspects of the propagation of
sound have been so far explored only in the linear regime
\cite{moelmer, javanainen, chiofalo, stoof, kraemer2002, rey,
machholm, taylor, smerzi2003a, menotti2003, kraemer2003a,martikainen}.
The main question addressed in this work is the strong interplay
between nonlinear effects and the periodicity of the external
potential.

We assume a fully coherent condensate described by an effective
one-dimensional Gross-Pitaevskii (GP) equation.  In the uniform case,
the spectrum of elementary excitations is given by the well-known
Bogoliubov dispersion relation $\omega(q)=\sqrt{q^2/2m \left( q^2/2m +
2U \right)}$, where $q$ is the momentum of the excitation, $m$ the
atomic mass and $U=gn$ the inverse compressibility with $g=4\pi\hbar^2
a/m$ the interaction strength and $n$ the density.  For momenta $q$
smaller than the inverse of the healing length $\xi=1/\sqrt{8 \pi a
n}$, the excitation spectrum depends linearly on the momentum with a
slope $c=\sqrt{U/m}$ which defines the sound velocity.

We now consider a condensate in a one dimensional optical lattice 
\begin{eqnarray}    
V_{opt}(x)= s\; E_R   \;{\rm sin}^2\left( {\pi x \over d} \right),    
\label{Vext}    
\end{eqnarray}    
characterized by a lattice spacing $d$ and a depth $s$ in units of the
recoil energy $E_R=\hbar^2 \pi^2/2md$. The elementary excitations of a
condensate in presence of a one dimensional lattice have been
discussed in detail, e.g., in \cite{kraemer2003a}. One of the main
features is the formation of a band structure in the Bogoliubov
spectrum when the lattice is turned on, in analogy with the linear
Bloch theory. The spectrum is periodic in the quasi-momentum $q$ of
the excitations and, for a given $q$, many excitation energies,
labeled by a band index, are available. Energy gaps open at the
boundary and at the center of the Brillouin zone, given respectively
by the Bragg momentum $q_B=\hbar\pi/d$ and $q_B=0$.

As in the uniform case, the energy of low energy excitations is linear
in the quasi-momentum, $\hbar\omega=c |q|$, and the system admits
sound waves with velocity
\begin{eqnarray}
c=\sqrt{\frac{U}{m^*}}.
\label{c}
\end{eqnarray}
The inverse compressibility $U$ increases very weakly with the optical
lattice depth $s$, and the effective mass $m^*$ instead strongly
increases with $s$, reflecting the corresponding quenching of the
tunneling rate through the barriers.  As a consequence the sound
velocity is predicted to decrease for increasing lattice depth.  In
the tight binding limit, the effective mass is related to the
tunneling parameter $\delta$ by $m^*/m=2 E_R /(\pi^2 \delta)$ and the
sound velocity can thus be written as $c= d \sqrt{U \delta} / \hbar$
\cite{smerzi2003a,kraemer2003a}.

However, it is not obvious whether a sound signal of observable
amplitude can propagate also in deep lattices, where the tunneling
rate is very small and nonlinear corrections to the Bogoliubov theory
can be important. For instance, nonlinear effects combined with the
presence of periodic trapping potentials drammatically affect the
center-of-mass motion of harmonically driven condensates due to the
onset of dynamical instabilities \cite{smerzi2002, menotti2003,
cataliotti2003}.

This paper is organized as follows: In Sect.~\ref{generation}, we
outline the procedure to generate a sound signals. In
Sect.~\ref{GPEDNLS} we introduce our theoretical analysis, based on
both the Gross-Pitaevskii (GP) equation and the discrete nonlinear
Schr\"odinger equation (DNLS).  In Sect.~\ref{vel} we obtain the
velocity of propagation of large amplitude wavepackets, and in
Sect.~\ref{nonlinearpropagation} we discuss the different regimes for
sound propagation. In Sect.~\ref{harmonictrapping} we comment on the
consequences of an additional harmonic trapping and, finally, in
Sect.~\ref{experimental} we conclude discussing the experimental
observability of the predicted effects.

\section{Generation of Sound Signals}
\label{generation}

\begin{center}
\begin{figure}
\includegraphics[width=0.9\linewidth]{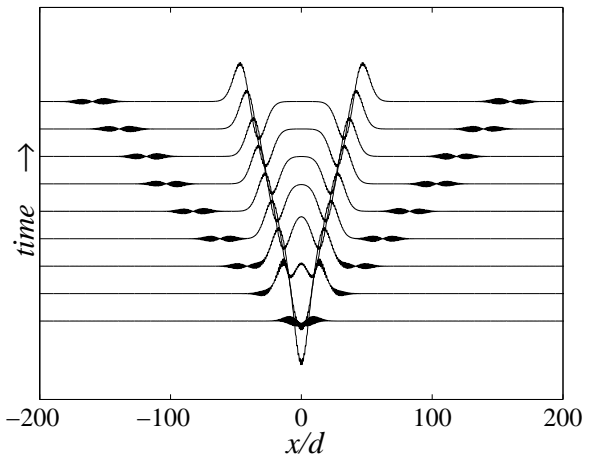}
\caption{GP simulation of wavepackets produced by a fast ($T_p=1$)
perturbation of the type (\ref{barrier}) in presence of an optical
lattice ($s=1$, $gn=0.5 E_R$).  Both the first, the second and the
third bands are excited leading to the formation of two pairs of
wavepackets.  For a slow perturbation ($T_p \gg 1$) only the slow pair
composed of phonons of the lowest band is created. Here, we plot the
relative density $n(x,t)/n(t,0)$ as defined in Sect.\ref{GPEDNLS}}
\label{higher_bands}
\end{figure}
\end{center}

One possibility to create excitations in the condensate consists in
turning on and/or off a perturbation potential in the center of the
system, for instance by means of a far detuned laser beam.  If the
width of the perturbing potential is much larger than the lattice
spacing, only quasi-momenta much smaller than the Bragg momentum $q_B$
are addressed. This procedure generates a pair of wavepackets
propagating symmetrically outwards.  Moreover, only Bogoliubov bands
with energy lower than the inverse time scale $T_p$ of the
perturbation are excited.  In particular, since the gap between first
and second Bogoliubov band at the center of the Brillouin zone is of
the order of $4E_R$, one has multiple-band excitations if $T_p < \hbar
/ 4E_R$, otherwise if $T_p \gg \hbar / 4E_R$ only the lowest band is
addressed.  The result of a fast perturbation ($T_p=\hbar/E_R$) at low
lattice depth is shown in Fig.\ref{higher_bands}. One finds two pairs
of wavepackets propagating at two different velocities. The velocity
of the slower ones is given by the sound velocity (\ref{c}), while the
velocity of the faster ones is given approximatively by $2 q_B/m$,
close to the derivative of the second and third band of the spectrum
at small $q$.  Those fast wavepackets are not sound waves but are
superpositions of single particle excitations: they disappear in
absence of the lattice, but if the lattice is present they are found
in a non-interacting gas as well.  Apparently, they travel without
changing shape, simply because the curvature of the spectrum is too
small to observe their dispersion on the time scale of our simulation.
This effect can be observed only in presence of shallow lattices, when
the gap between second and third band is negligible. For larger $s$,
such single particle excitations travel with smaller velocity, disperd
more quickly, and can interfere with the propagation of the sound
wavepackets. For this reason it is better to use slower perturbations
$T_p \gg 1$ in order to restrict the dynamics to the first band.

The shape of the wavepackets depends on the excitation 
producedure. However, the main features of their behaviour are quite
general. In absence of the lattice, the observation of sound signals
in a harmonically trapped condensate was achieved experimentally by
employing two different excitations methods \cite{andrews1997a}:
raising a potential barrier in the center of the trap produces a
density bump, which splits into two {\it bright} sound wavepackets;
alternatively, removing a potential barrier from the center of the
condensate gives rise to a dip in the density which splits into two
{\it grey} sound wavepackets.

The excitation method we adopt here is a combination of these two: The
initial condensate is in the ground state of a one-dimensional optical
lattice. We then switch on and off a gaussian potential in the center
and get two composed {\it bright-grey} sound signals propagating
symmetrically outwards.  This procedure has the advantage that the
ground states of the initial and final potential are identical.  

The perturbing potential has a spatial and temporal dependence
\begin{equation}
V_P(x,t)=V_{Px}(x) V_{Pt}(t)\,,
\label{barrier}
\end{equation}
where 
\begin{eqnarray}
V_{Px}(x)&=&b E_R \exp\left[-x^2/(wd)^2\right]\,,
\label{barrier_spatial}\\
V_{Pt}(t)&=& \left\{
\begin{array}{c l}
0, & \; {\rm for \;\;} t<0, \\
\sin^4\left({\pi E_R t \over \hbar T_p}\right)\,, 
  & \; {\rm for \;\;}  0<t<T_p\hbar/E_R \\
0, & \; {\rm for \;\;} t> T_p\hbar/E_R .
\end{array} 
\right.
\label{barrier_switch}
\end{eqnarray}
The tunable dimensionless parameters are the width of the potential
$w$, its height $b$ and the time of perturbation $T_p$.  
We impose the constraints
\begin{equation}
w \gg 1\,,
\label{w}
\end{equation}
in order to address only the quasi-momenta in the central part of the
Brillouin zone, and
\begin{equation}
T_p >1
\end{equation}
in order to excite the lowest Bogoliubov band only.  Note that for
typical densities, scattering lengths and lattice spacings, $\xi$ is
of the order of $d$. Hence, $w\gg 1$ automatically implies $wd \gg
\xi$, which ensures that the produced excitations are phonons.


\section{Gross-Pitaevskii (GP) and Discrete Nonlinear 
Schr\"odinger Equations (DNLS)}
\label{GPEDNLS}

We study the sound propagation in the presence of the lattice
potential (\ref{Vext}) with the one-dimensional GP-equation
\begin{eqnarray}
i \hbar {\dot \varphi} = 
\left[ 
- \frac{\hbar^2 \partial_x^2}{2m} + V(x,t) + gnd |\varphi(x,t)|^2 
\right]
\varphi(x,t)\, ,
\label{gpe_sound}
\end{eqnarray}
describing a system uniform in the transverse directions $y,z$ with
$n$ the 3D average density. The consequences of a transverse confinement,
which can modify the degree of nonlinearity of the system and the
speed of sound \cite{smerzi2003a}, are discussed in
Sect.\ref{harmonictrapping}. The normalization of the wavefunction is
$\int_{-N_w d/2}^{N_w d/2} |\varphi(x,t)|^2 dx= N_w$, with $N_w$ the
total number of lattice wells.  The external potential $V(x,t)$ is
given by the sum of the lattice potential $V_{opt}=s E_R\sin^2(\pi
x/d)$ and the time-dependent perturbation $V_P(x,t)$ written in
Eq.(\ref{barrier})
\begin{eqnarray}
V(x,t)=V_{opt}(x)+V_P(x,t)\,.
\end{eqnarray}
The ground state of the system in the presence of the optical
potential is governed by two parameters: the optical lattice depth $s$
and the interaction strength $gn$.

Writing $\varphi(x,t)=\sqrt{n(x,t)/n} \exp [i S(x,t)]$, the
GP-equation (\ref{gpe_sound}) can be recast in the form of two coupled
equations for the density and phase variables
\begin{eqnarray}
\label{GPE2}
{\dot n}(x) &=& 
-\partial_x \left[ n(x) \frac{\hbar}{m} \partial_x S(x) \right] , \\
{\dot S}(x) &=& - \frac{1}{\hbar}
\left[ V(x,t) + gn(x) + \right. \nonumber \\
&& \left. - \frac{\hbar^2}{2m\sqrt{n}} \partial_x^2 \sqrt{n}
+ \frac{\hbar^2}{2m} \left(\partial_x S(x)\right)^2 \right].
\nonumber
\end{eqnarray}

\begin{center}
\begin{figure}[h]
\includegraphics[width=0.85\linewidth]{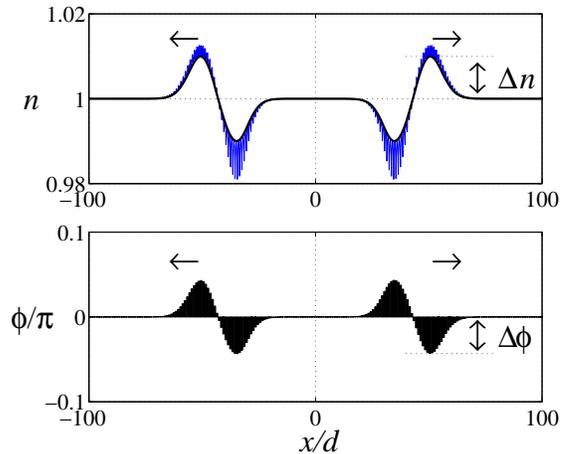}
\caption{ Typical result of a GP simulation for small perturbations
(linear regime).  Two sound packets move simmetrically outward at the
Bogoliubov sound velocity.  The relative density $n(x,t)/n(x,0)$ (thin
line), the convoluted signal (thick line) and the phase difference
$\phi_{\ell +1/2}$ are shown.  The signal amplitudes of the density
and relative phase are denoted by $\Delta n$ and $\Delta \phi$
respectively.  }
\label{n_ph}
\end{figure}
\end{center}

In the presence of the lattice, both the density and the phase are
characterised by very strong modulations on the length scale of the
lattice spacing $d$.  In order to highlight the density variation
corresponding to the sound wavepacket, it is in general convenient to
plot the density relative to the ground state density
$n(x,t)/n(x,0)$. Alternatively, in order to mimic the limited
resolution of a detection system, one can perform a convolution over a
few lattice wells of the GP results for $n(x,t)$.  We also consider
the evolution of the relative phase between neighboring wells, defined
as
\begin{equation}
\phi_{\ell+1/2}=S(x=(\ell+1)d)-S(x=\ell d).
\label{phi_gpe}
\end{equation}
The typical signals obtained in a GP simulation in the linear regime
are shown in Fig.(\ref{n_ph}), where in the upper planel we plot both
the relative density $n(x,t)/n(x,0)$ and the convoluted signal. We
denote by $\Delta n$ and $\Delta \phi$ the signal amplitudes of the
density and relative phase wavepackets respectively.

In the tight binding regime, where the condensate phase is
approximately flat in each well, the GP dynamics can be simplified by
using the discretised wavefunction
$\psi_{\ell}=\sqrt{n_{\ell}(t)}\exp[i S_{\ell}(t)]$, where
$n_{\ell}(t) = | \psi_{\ell}(t) |^2 = \int_{\ell d -d/2}^{\ell d +d/2}
| \varphi(x,t)|^2 dx$ and $S_{\ell}(t) = S(x=\ell d,t)$. At
equilibrium $n_{\ell}=1$ and $S_{\ell}=0$. The time evolution of
$\psi_{\ell}$ is given by the discrete nonlinear Schr\"odinger
equation (DNLS) \cite{trombettoni2001a}
\begin{eqnarray}
i \hbar {\dot \psi}_{\ell} = 
-\frac{\delta}{2} \left( \psi_{\ell +1} +  \psi_{\ell -1} \right)
+ \left[ V_{\ell}(t) + U |\psi_{\ell}(t)|^2 \right] \psi_{\ell}(t). \nonumber \\
\label{dnls_sound}
\end{eqnarray}
Here, the external potential $V_{\ell}(t)$ includes only the
perturbation $V_P(ld,t)$, with $V_P$ given by (\ref{barrier}), since
the presence of the optical lattice potential is accounted for by the
discretization of space.  The two quantities $\delta$ and $U$ entering
the DNLS describe, respectively, the tunneling coupling between two
neighbouring wells (corresponding to half the height of the lowest
Bloch band) and the on-site interaction $U$ (corresponding to the
inverse compressibility).

In terms of density and phase variables, the DNLS equation becomes
\begin{eqnarray}
\label{DNLS2}
{\dot n}_{\ell} &=& \sum_{\ell'=\ell \pm 1}
\frac{\delta}{\hbar} \sqrt{n_{\ell}(t) n_{\ell'}(t)}
\sin[S_{\ell}(t)-S_{\ell'}(t)] , \\
{\dot S}_{\ell} &=& -\frac{\mu_{\ell}}{\hbar}+
\sum_{\ell'=\ell \pm 1}
\frac{\delta}{2 \hbar} \sqrt{\frac{n_{\ell'}(t)}{ n_{\ell}(t)}}
\cos[S_{\ell}(t)-S_{\ell'}(t)]\,,
\nonumber
\end{eqnarray}
where $\mu_{\ell}=U n_{\ell} +V_{\ell}$.  

The results of the numerical integration of the DNLS equation
(\ref{dnls_sound}) match those obtained by solving the GP-equation
(\ref{gpe_sound}) in deep lattices.  In order to compare the density
$n(x,t)$ calculated with the GP equation with the solution
$n_{\ell}(t)$ of the DNLS equation, it is necessary to average
$n(x,t)$ over each lattice site.  Instead the comparison for the
evolution of the relative phase $\phi_{\ell+1/2}=S_{\ell+1}-S_{\ell}$
is strighforward.

The DNLS approach is not only convenient from the numerical point of
view, but also allows a clearer and semi-analitycal understanding of
the basic physics underlying sound propagation and shock-waves
formation in the tight binding regime.  In particular the DNLS
equation written in the form (\ref{DNLS2}) is particularly useful to
get insight into the breakdown of the linear regime for deep lattices
(see discussion in Sect.\ref{nonlinearpropagation}).

In the following, we will discuss the solution of the GP-equation
(\ref{gpe_sound}) and DNLS equation (\ref{dnls_sound}) for various
values of the potential height $b$, the potential width $w$ and the
perturbation time $T_p$.  We will vary the parameters $s$ and $gn$ in
the case of the GP equation and the parameters $\delta$ and $U$ in the
case of the DNLS to explore different regimes of lattice depths and
interaction strengths.

\section{Velocity of propagation of wavepackets}
\label{vel}

The first quantity we want to extract from our simulations is the
velocity of propagation of sound wavepackets as a function of lattice
depth. The result in the linear regime has already been predicted in
\cite{kraemer2003a}.  Here we focus on the case of large perturbations
producing signals which can be observed experimentally.

\begin{center}\begin{figure}[h!]
\includegraphics[width=0.85\linewidth]{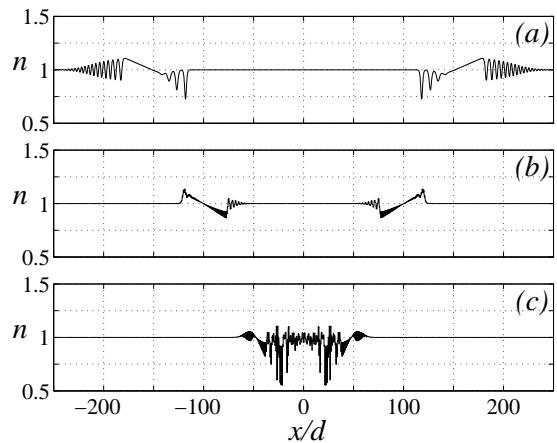}
\caption{GP simulation for the relative density $n(x,t)/n(x,0)$ at
$t=480 \hbar/E_R$ with $gn=0.5 E_R$ for different lattice depths:
$s=0$ (a), $s=10$ (b) and $s=20$ (c).}
\label{fig_s0_sinter_slarge}
\end{figure}
\end{center}

\begin{center}
\begin{figure}
\includegraphics[width=0.85\linewidth]{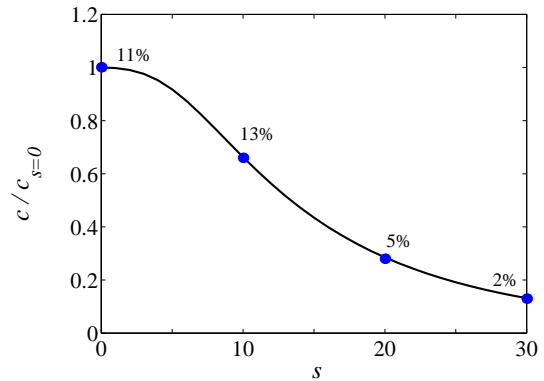}
\caption{ Sound velocity as a function of lattice depth $s$.
Bogoliubov prediction (solid line) and results ``measured'' based on
the simulation (circles) with respective signal amplitudes $\Delta n$
for $gn=0.5E_R$.  The signal amplitude $\Delta n$ is defined as
indicated in Fig. \ref{n_ph}.  }
\label{fig_soundpropagation}
\end{figure}
\end{center}

In Fig.~\ref{fig_s0_sinter_slarge} we plot some examples for the
relative density at the final stage of our GP-simulation at lattice
depths $s=0,\,10,\,20$ with $gn=0.5 E_R$ for a large perturbation.
Those signals involve significant nonlinear effects, even in the
uniform case (see Fig.\ref{fig_s0_sinter_slarge}(a)).  From these
simulations, after performing the convolution of the signal, we obtain
the sound velocity shown by the circles in
Fig. \ref{fig_soundpropagation}. The respective signal amplitudes
$\Delta n$, together with the Bogoliubov prediction (solid line) are
also indicated in the figure. 

The first conclusion is that from our simulation we can extract the
value of the Bogoliubov sound velocity with high accuracy. Moreover,
we find that sound signals of measurable amplitude can be observed
also in deep lattices where the sound velocity is considerably lower
than in the uniform system.  At this value of $gn$, relatively large
signal amplitudes can be obtained up to $s=20$.  Note that the signal
amplitudes obtained from the simulation at $s=20,\,30$ correspond to
their saturated values as discussed in detail further below.

\section{Linear and nonlinear regimes} 
\label{nonlinearpropagation}

By keeping lattice depth and interaction fixed while increasing the
strength of the external perturbation, the role of nonlinearities
becomes more and more important.  The system passes through three
regimes:
\begin{enumerate}
\item 
linear regime, where the Bogoliubov description holds and the
variations of density and relative phases are small;
\item 
shock wave regime, where density variations induce mode--coupling
among Bogoliubov excitations, giving rise to the formation of shock
waves.  Depending on the curvature of the Bogoliubov dispersion beyond
the phononic regime, shock waves emerge in front of the wavepacket
(uniform system, shallow lattice), both in the front and in the back
(intermediate lattice) or only in the back (deep lattice with
$\delta\ll U/3$).  In the first two cases, the sound signal is deformed
and it is finally disperded. 
Instead, when shock waves form only in the
back, the signal maintains a compact shape and propagates at the sound
velocity predicted by Bogoliubov theory.
\item 
Saturation regime, where the sound signal amplitude saturates. The
sound wavepacket leaves behind a wake of noise, and still propagates
at the sound velocity predicted by Bogoliubov theory.  This regime
exists only in the presence of the lattice and provided that
$\delta\ll U/3$ to ensure that shock waves form in the back of the
signal as explained in regime (2). For fixed $gn$ or $U$, the
condition $\delta\ll U/3$ can be ensured by making the lattice
sufficiently deep.
\end{enumerate}
The signal amplitudes attainable in each regime and the perturbation
parameters $b,\,w,\,T_p$ needed to reach a certain regime depend on
the lattice depth $s$ and on the interaction strength $gn$, or
equivalently, on $\delta$ and $U$.  In the presence of a lattice, a
stronger perturbation is needed to obtain the same signal amplitude as
in absence of lattice. This reflects the fact that the condensate in a
lattice is less compressible.  In the following subsections we are
going to discuss more in detail the three regimes.

\subsubsection{Linear Regime}

In the linear regime, the signal moves without dispersion and with
sound velocity predicted by the Bogoliubov theory.  The amplitudes
$\Delta n$ and $\Delta\phi$ are small and remain constant during the
propagation.  The ratio between the relative density and the relative
phase amplitudes can be derived analytically.  This can be done using
the DNLS equations (\ref{DNLS2}) valid only for deep lattices, or the
macroscopic hydrodynamic formalism developed in \cite{kraemer2003a}
valid for all lattice depths and small quasimomentum.

A solution is provided by the ansatz 
\begin{eqnarray}
{\bar n}(x,t) &=&  1+\Delta n [ f_+(x-ct)+f_-(x+ct)], \\
\phi(x,t) &=& -\Delta \phi [ f_+(x-ct)-f_-(x+ct)],
\end{eqnarray}
where ${\bar n}$ represent either the DNLS density or the macroscopic
averaged density.  The functions $f_+$ and $f_-$ describe respectively
the wavepackets moving to the left and to the right, and $\Delta n$
and $\Delta\phi$ are their amplitudes, as shown in Fig.\ref{n_ph}.  We
obtain
\begin{eqnarray}
\Delta n = \frac{\hbar}{d} \sqrt{\frac{1}{U m^*}} \Delta \phi 
\;\;\;\;\;\;\;\longrightarrow \;\;\, 
\hspace*{-1.1cm} \raisebox{0.25cm}{{\rm DNLS}} \;\;\;
\sqrt{\frac{\delta}{U}} \Delta \phi,
\label{deltan_deltaphi}
\end{eqnarray}
where the last expression is written in terms of DNLS parameters.  We
notice that in order to get the DNLS relation, we have replaced
$\sin[S_{\ell}(t)-S_{\ell'}(t)] $ with its argument, which implies
$S_{\ell}(t)-S_{\ell'}(t)\ll\pi$. When this condition is not
satisfied, non linear effects due to the optical lattice becomes
important, and the linearised formalism breaks down.

\subsubsection{Shock Wave Regime}

The peculiarity of this region is the formation of shock waves.  In
the uniform case, a front wave emits shock waves in the forward
direction (see Fig.\ref{fig_s0_sinter_slarge}(a)).  The stronger is
the external perturbation, the stronger is the deformation and
spreading of the sound wavepacket due to the emission of shock
waves. We smoothen out the large density fluctuations in the shock
wave region with a gaussain convolution of the signal over few lattice
sites, mimicking the limited resolution of a detection system.  A
measurement of the position of maximum or the minimum of the resulting
signal, propagating, respectively, faster and slower than sound,
yields a velocity which deviates from the Bogoliubov prediction.
However, in spite of the deformation of the signal, it is possible to
extract the Bogoliubov value of the sound velocity as shown in
Fig.\ref{fig_soundpropagation}, by following the center-of-mass
position of the signal \cite{note2}.

In a shallow lattice, shock waves form in the front as in the uniform
case, because the formation of a gap in the Bogoliubov spectrum does
affect only a small range of quasi-momenta close to the Brillouin zone
boundary. Hence, the mode--coupling among Bogoliubov excitations leads
to the creation of excitations outside the phononic regime which
travel at a speed larger than the sound velocity.

\begin{center}
\begin{figure}
\includegraphics[width=0.75\linewidth]{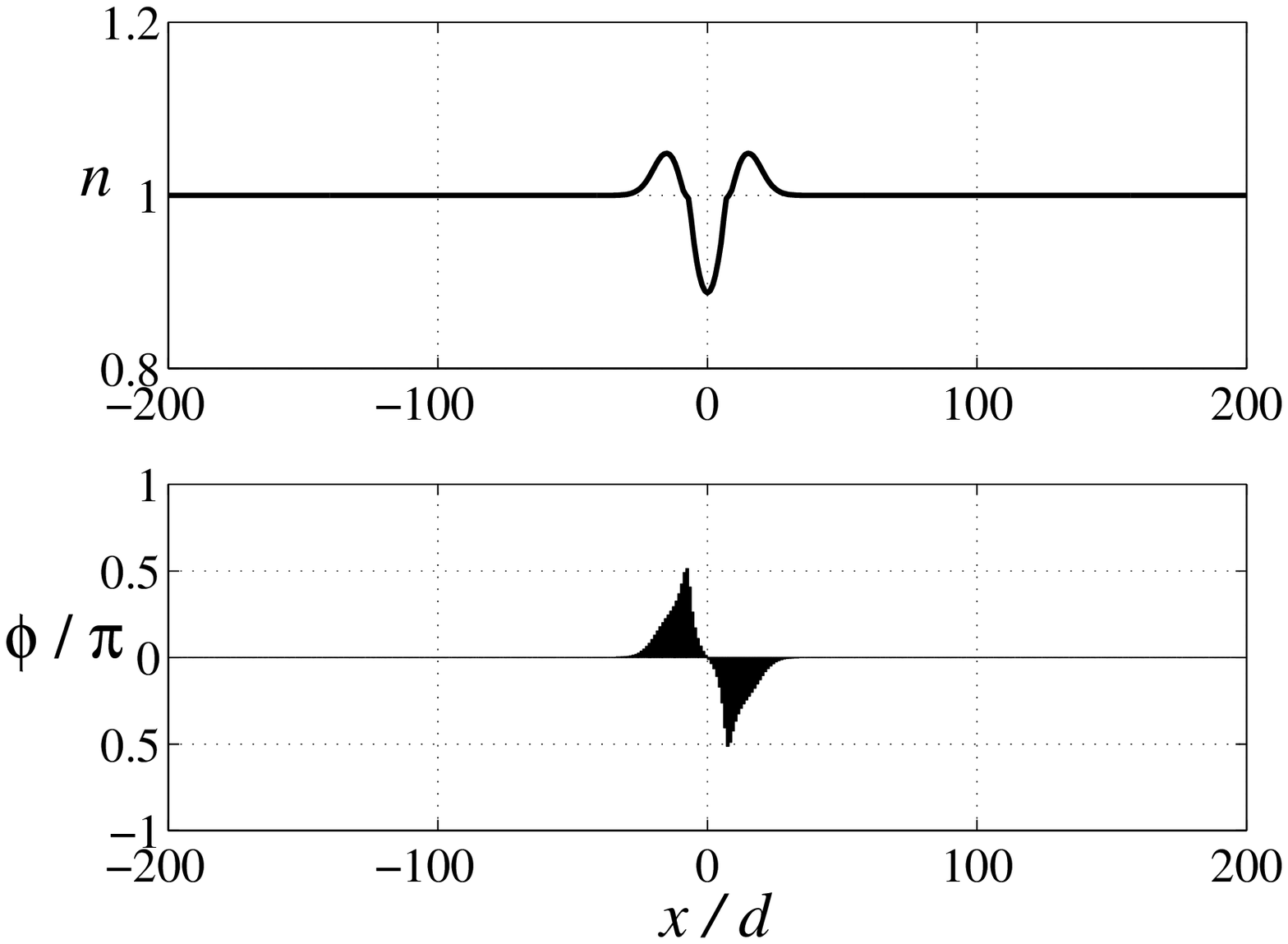}
\includegraphics[width=0.75\linewidth]{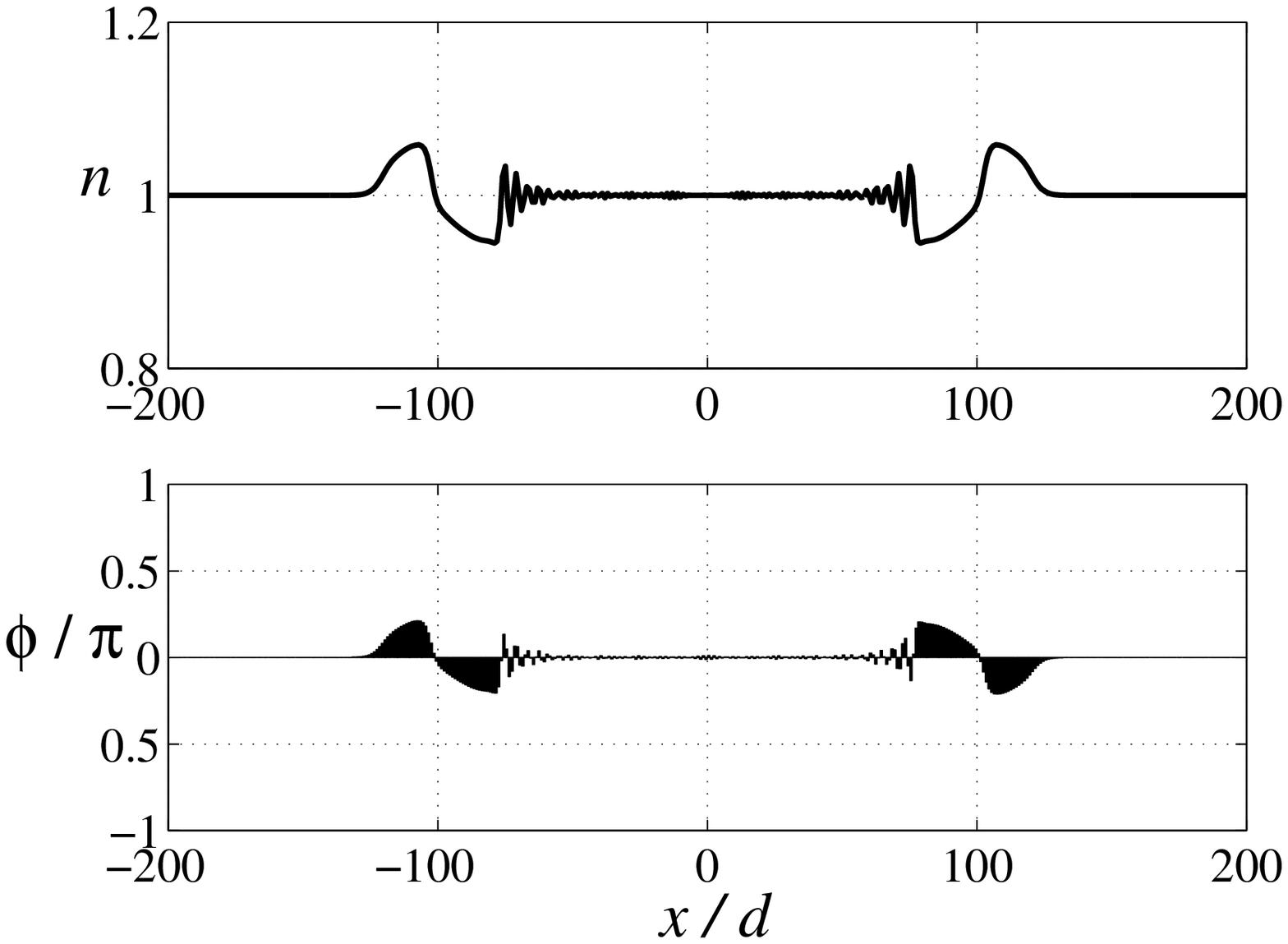}
\caption{DNLS simulation of the density $n{\ell}(t)$ and relative
phase $\phi_{\ell +1/2}$ for a deep lattice ($\delta/U = 10^{-2}$) in
the shock wave regime ($T_pb/w \sim 5.5$). Upper panels: early stage
($t=16\hbar/E_R$); lower panels: late stage ($t=200\hbar/E_R$).}
\label{fig_shockwave_lattice}
\end{figure}
\end{center}

On the other hand, in intermediate and deep lattices shock waves can
also form behind the sound packet due to the behaviour of the lowest
band Bogoliubov dispersion, which in the tight binding regime takes
the form \cite{javanainen,kraemer2003a}
\begin{eqnarray}    
\label{bog_tb}    
\hbar \omega_q \approx 
\sqrt{ 2\delta{\rm sin}^2\left({\pi q \over 2 q_B }\right) 
\left[ 2 \delta
{\rm sin}^2 \left({\pi q \over 2 q_B}\right) + 2 U \right]}.
\end{eqnarray}    
For typical values of the density, the ratio $\delta/U$ lies between
zero and one. If the ratio $\delta/U$ is larger than $1/3$, the
Bogoliubov dispersion has a positive curvature in a small range of
quasimomenta and becomes negative closer to the zone boundary. In this
case the shock waves are produced both in the front and in the back of
the sound wavepacket. The dispersion has a negative curvature for all
$q$ as long as $\delta/U<1/3$. In deep lattices (where $\delta/U \ll
1/3$), wavepackets composed by quasi-momenta outside the phononic
regime will propagate much slower than sound and shock waves are
therefore created behind the sound wavepacket
(Fig.\ref{fig_s0_sinter_slarge}(b) and
Fig.\ref{fig_shockwave_lattice}).  In our simulations, for $\delta / U
= 10^{-2}$, we observe that the relative phase distribution is
strongly deformed if $\phi_{\ell+1/2} \sim \pi/2$, due to the non
trivial {\rm sin}-dependence of the current in the DNLS
Eq.(\ref{DNLS2}) (see Fig. \ref{fig_shockwave_lattice} (upper two
panels)).  However this behaviour does not become critical up to the
point where the relative phase of two neighboring lattice sites
$\phi_{\ell+1/2} = \pi$, which defines the onset of regime (3).

\subsubsection{Saturation Regime in deep lattices}
\label{satdeep}

\begin{center}
\begin{figure}
\includegraphics[width=0.75\linewidth]{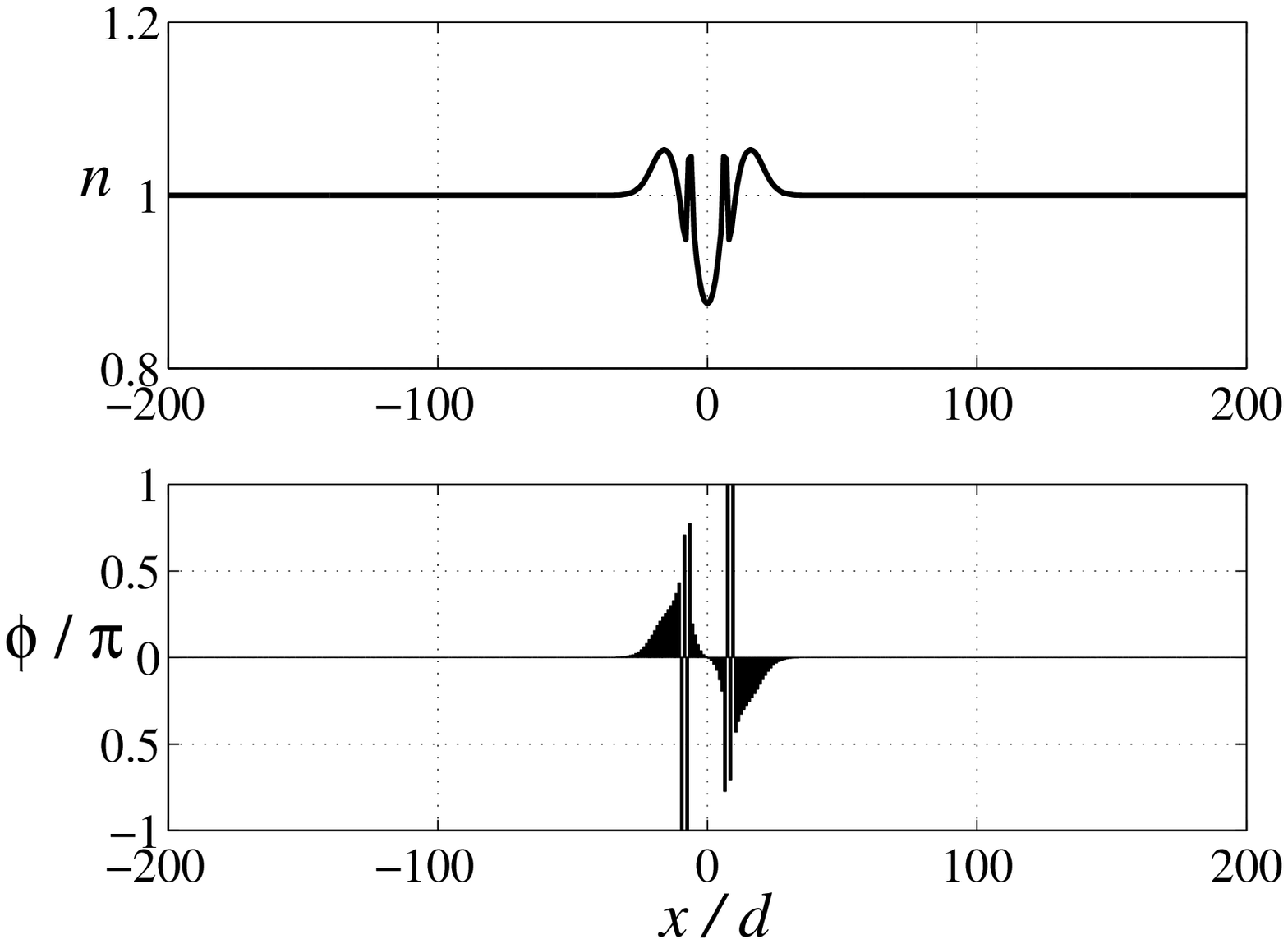}
\includegraphics[width=0.75\linewidth]{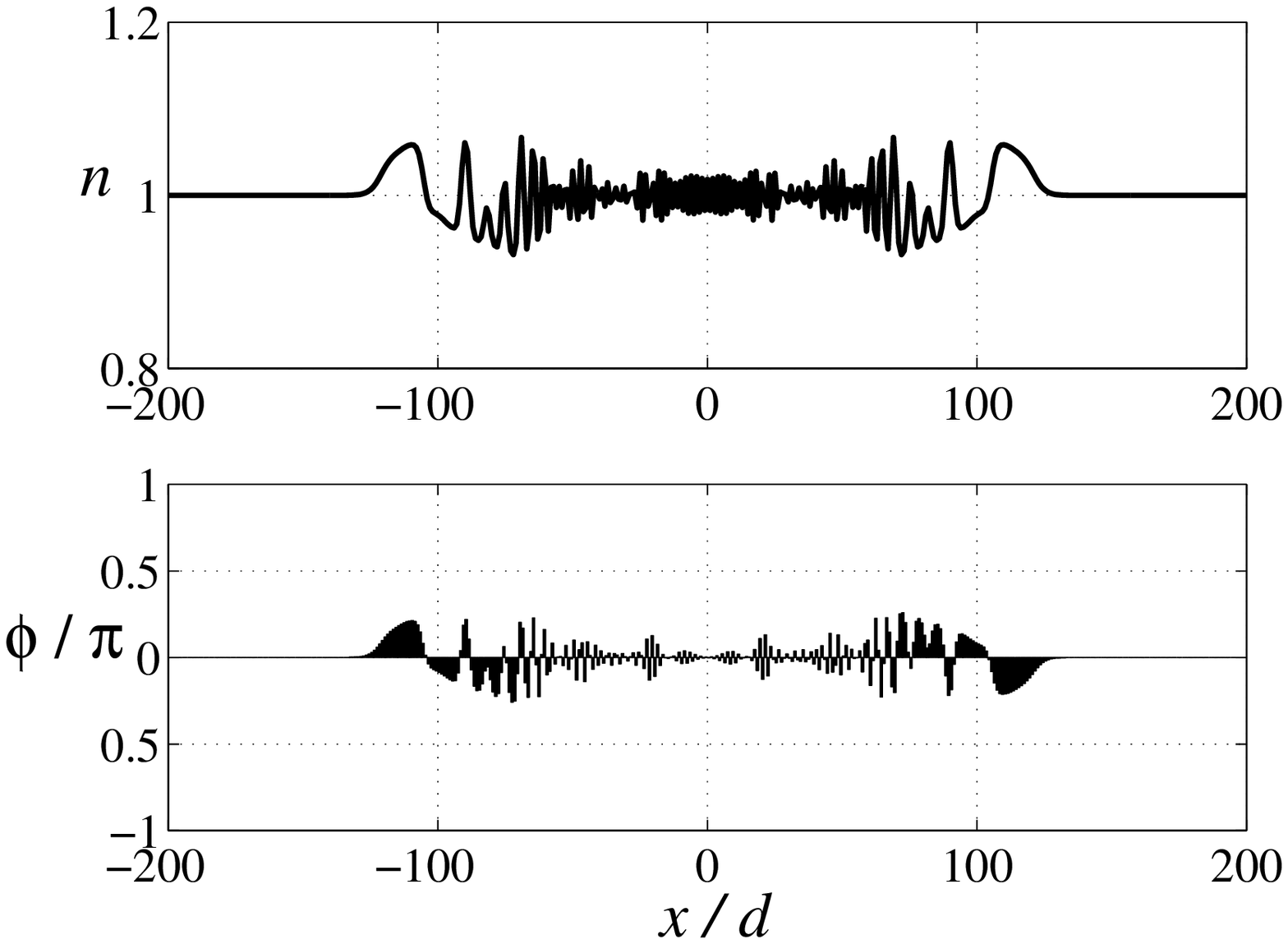}
\caption{DNLS simulation of the density $n_{\ell}(t)$ and relative
phase $\phi_{\ell +1/2}$ for a deep lattice ($\delta/U = 10^{-2}$) in
the saturation regime ($T_pb/w \sim 7$). Upper panels: early stage
($t=16\hbar/E_R$); lower panels: late stage ($t=200\hbar/E_R$).}
\label{bh50}
\end{figure}
\end{center}

If we further increase the strength of the external perturbation, the
relative phase $\phi_{\ell+1/2}$ at some site reaches $\pi$ (see
Fig.\ref{bh50}, upper two panels) and there the current starts flowing
in the opposite direction (see Eq.(\ref{DNLS2})).  As a consequence, a
wake of noise is left behind the sound packets and we find a
saturation in the amplitude of the propagating signal (see
Fig.\ref{fig_s0_sinter_slarge}(c) and Fig.\ref{bh50}, lower two
panels). The interesting feature is that the noise has zero average
velocity, since the oscillations of population between different wells
are completely dephased. Provided that the lattice is deep enough to
satisfy $\delta\ll U/3$, shock waves form only in the back of the
signal, as discussed above. Hence the noise and the shock waves never
overtake the sound signal, which is always able to ``escape'' from
them.

We stress that this effect appears only in presence of a deep
lattice. In fact, in the uniform case, even in presence of strong
nonlinearities leading to a strong deformation of the signal, in the
central region the system is always able to recover its ground state
after the sound wave has passed by (see Fig.\ref{fig_s0_sinter_slarge}
(a)).

\begin{center}
\begin{figure}
\includegraphics[width=0.75\linewidth]{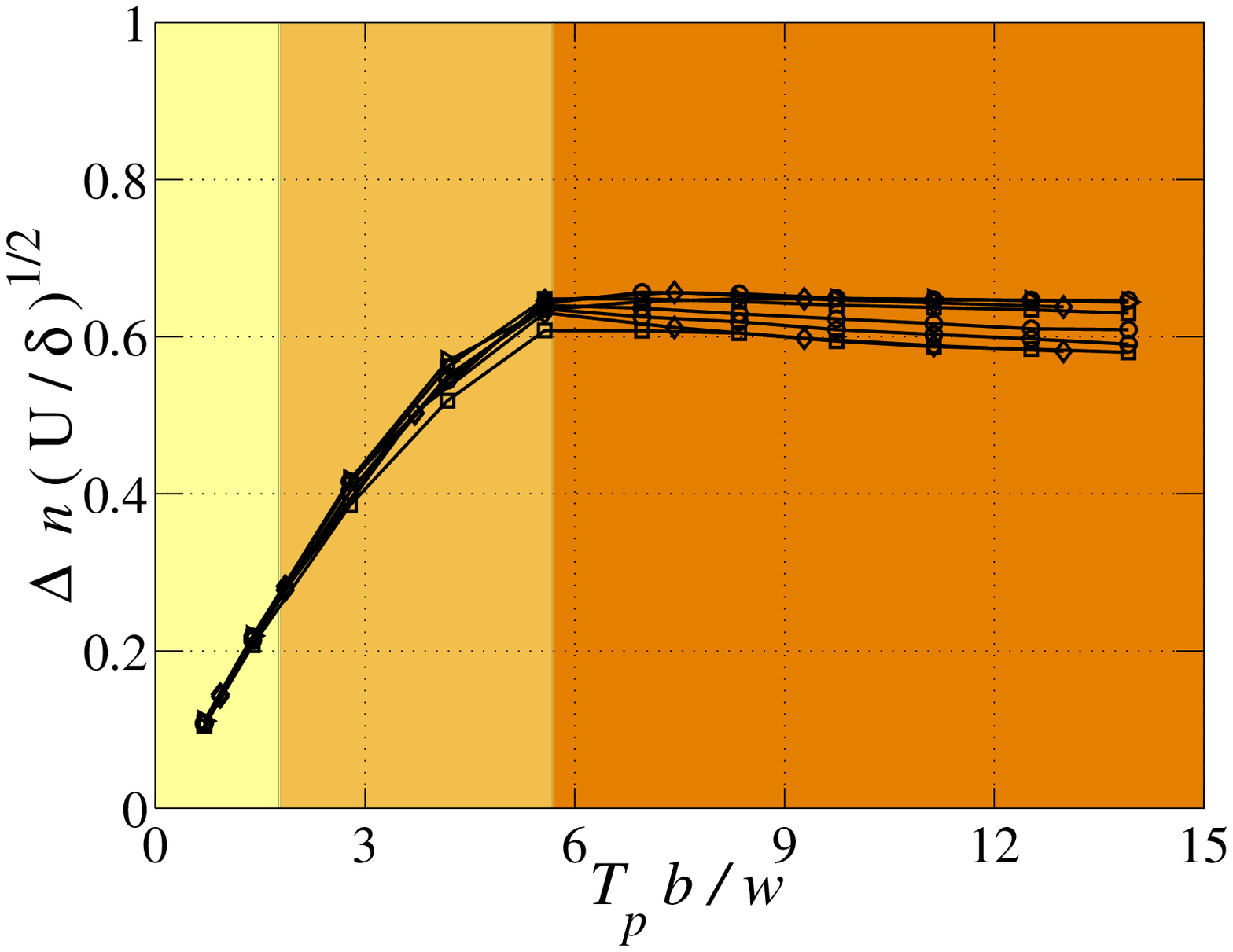}
\includegraphics[width=0.75\linewidth]{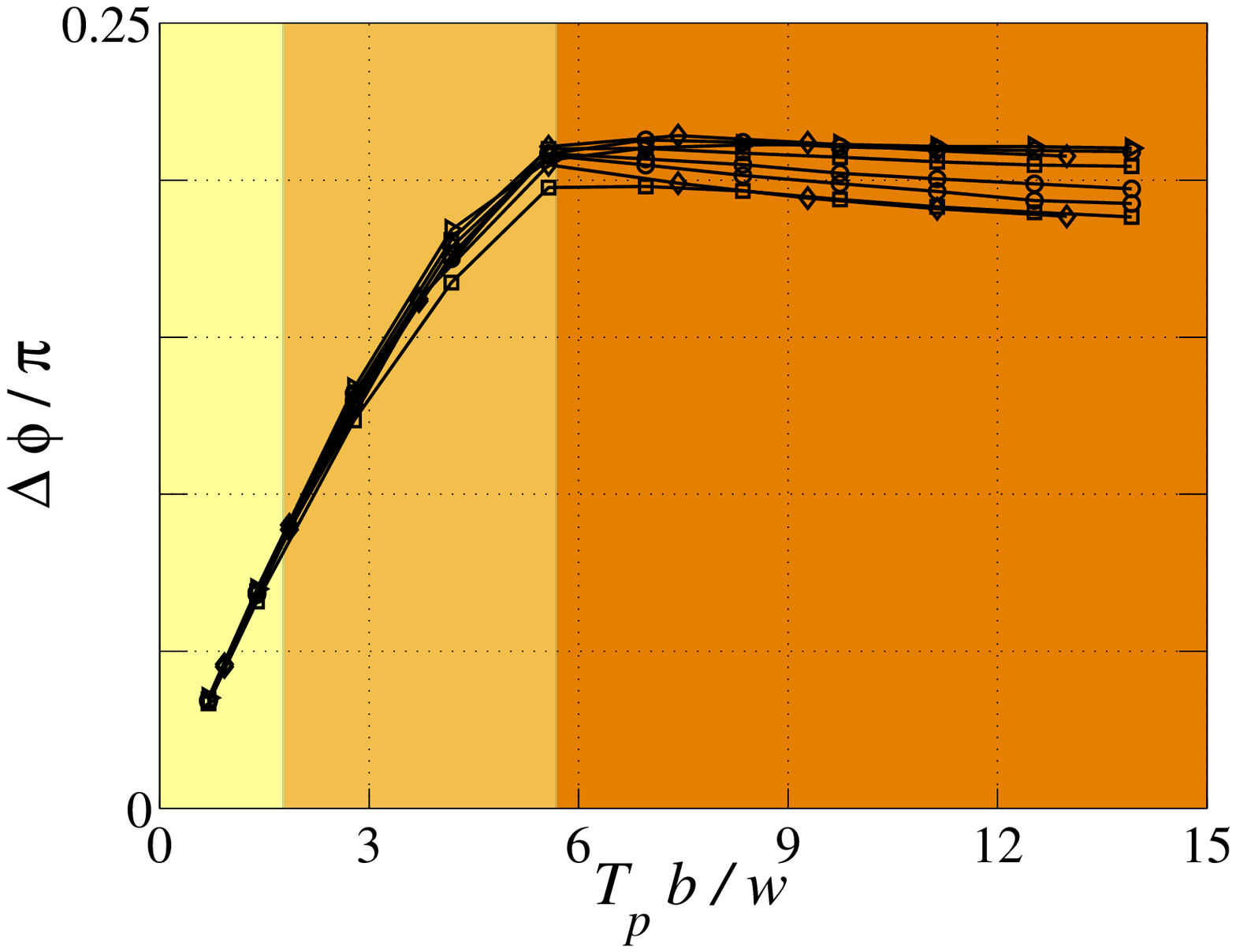}
\caption{Results for the density and relative signal amplitudes
$\Delta n$ and $\Delta \phi$ obtained from DNLS simulations for
varying perturbation parameters $T_p,\,b,\,w$ (see Eq.(\ref{barrier})),
tunnel coupling $\delta$ and on-site interaction $U$ (see
Eq.(\ref{dnls_sound})) with $\delta\ll U/3$.  }
\label{post_sl}
\end{figure}
\end{center}

To demonstrate the saturation of the signal, we follow the evolution
of the system for a long time and look at the amplitudes $\Delta n$
and $\Delta\phi$.  Both in the GPE (for relatively deep lattices) and
in the DNLS simulations, we find an interesting scaling law that helps
to distinguish between the three regimes mentioned above and makes
evident the saturation of the signal amplitude.  The scaling of the
results is shown in Fig.\ref{post_sl}: The effect of the perturbing
potential takes a universal form when:

\begin{itemize}
\item 
the perturbation parameters are combined in the form $T_p b /w$ which
reflects the capability of the system to react to an external
perturbation;
\item 
the relative density variation is rescaled as $\Delta n \sqrt{U
/\delta}$, while the amplitude of the relative phase signal $\Delta
\phi$ does not need to be rescaled.
\end{itemize}

The results summarized in Fig.\ref{post_sl}, show that for small
parameter $T_p b /w$, the perturbation produced in the system is
small, and depends linearly on $T_p b /w$.  Increasing the
perturbation parameter $T_p b /w$, the signal amplitude $\Delta n$
quickly saturates. The three regions indicated in Fig.\ref{post_sl}
correspond to the three regimes (1-3) listed above.

From our numerical results, we find that the proportionality
between the relative density and the relative phase amplitudes written
in Eq.(\ref{deltan_deltaphi}), holds also beyond the linear regime. In
particular, we found a saturation of the relative phase amplitude at a
value $\Delta \phi \approx 0.2\pi$, which goes with the saturation of
the density amplitude $\Delta n$.  This implies that for given
interaction and lattice depth, the amplitude of the density signal is
limited by approximately $\Delta n \approx 0.2\pi\sqrt{\delta/U}$.

\begin{center}
\begin{figure}
\includegraphics[width=0.75\linewidth]{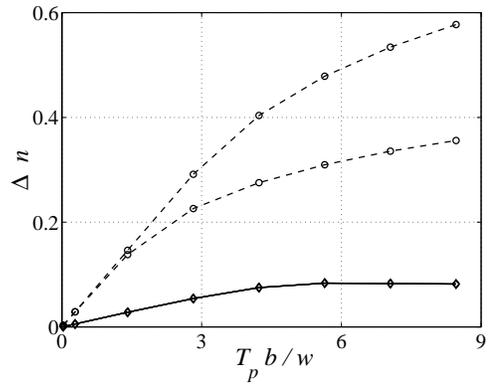}
\caption{ The signal amplitude $\Delta n$ as a function of the
perturbation parameter $b\,T_p/w$ at $gn=0.5E_R$.  Dashed lines: GP
simulation for $s=0$ and $gn=0.5 E_R$ at $t=100\hbar/E_R$ (upper
dashed line) and $t=200\hbar/E_R$ (lower dashed line).  Solid lines:
GP simulation for $s=15$ and $gn=0.5 E_R$ at
$t=100,\,200,\,300,\,400\hbar/E_R$ for $s=15$. The corresponding four
lines exactly overlap.  }
\label{fig_saturation1}
\end{figure}
\end{center}

Saturation does not occur in the uniform system or at low lattice
depth. To demonstrate this, we plot in Fig. \ref{fig_saturation1} the
signal amplitude $\Delta n$ measured after different propagation times
for $s=0$ (dashed lines) and $s=15$ (solid lines) as a function of
potential height $b$, with fixed potential width $w$ and perturbation
time $T_p$. With $s=15$ the signal amplitude takes exactly the same
values at $t=100,\,200,\,300,\,400\hbar/E_R$ (the corresponding four
lines in Fig. \ref{fig_saturation1} perfectly overlap!). In contrast,
the signal amplitudes in the uniform system ($s=0$, dashed lines) at
different times ($t=100,\,200\hbar/E_R$) differ from each other: At
earlier times the signal amplitude increases as a function of the
strength of the external perturbation and does not saturate. The
amplitude measured at later times coincides with the one measured
earlier only for small potential heights $b$, when the propagation
dynamics is linear. When nonlinear effects becomes important, $\Delta
n$ decreases during the evolution as a consequence of the dispersion
of the signal.

\section{Role of harmonic trapping}
\label{harmonictrapping}

Realistic experimental setups include an harmonic trapping
potential in the parallel and transverse directions
of the lattice.

In the longitudinal direction, the harmonic trap has a negligible
effect on sound propagation if $L \gg wd$, being $L$ the size of the
condensate and $w$ the width of the perturbing potential. This
condition guarantees, first of all, that the sound wavepacket travels
in a region of approximatively constant average density.  Moreover it
ensures that low energy discretised modes \cite{stringari1998a} are
not strongly excited.  

In the Thomas-Fermi regime, we find that the time required by the
sound signal to travel along a distance of the order of the size of
the condensate is $\sim 2 \pi / \omega_D$, with $\omega_D=\sqrt{m/m^*}
\omega_x$ the renormalised dipole frequency
\cite{kraemer2002,cataliotti2001}.

The transverse harmonic trapping leads to an inhomogeneous radial
density profile which can affect the sound velocity. Even in the
uniform case, a radial Thomas-Fermi density distribution changes the
sound velocity $c=\sqrt{U/m}$ to $c=\sqrt{U/2m}$
\cite{andrews1998a,zaremba1997a,kavoulakis1997a,stringari1998a}.  The
analogous modification of the velocity of sound to $c=\sqrt{U/2m^*}$
is expected in presence of the lattice provided the inverse
compressibility $U$ is linear in the density, as discussed in
\cite{smerzi2003a,kraemer2003a,martikainen}.  This approximation
requires the interactions $gn$ to be sufficiently small and has proven
applicable to describe the experimental result on collective
excitations \cite{fort2003a} and on the condensate size
\cite{morsch2002} where $gn \approx 0.2 \div 0.5 E_R$.

Moreover the predicted effects require single band dynamics.  This is
ensured if the radial trapping frequency is larger than the width of
the first Bogoliubov band. In the tight binding regime, this condition
reads $\hbar \omega_{\perp} > 2 \sqrt{U \delta}$.

\section{Conclusions}
\label{experimental} 

In conclusion, we find that there exists a broad range of optical
potential depths where the change in sound velocity induced by the
lattice should be measurable also in the non linear regime.  For
instance, a velocity of propagation $c=0.65 c_{s=0}$ and $0.30
c_{s=0}$ can be measured at an optical lattice depth of $s=10$ and
$20$ with a maximal density variation $\Delta n_{\rm max}=0.13$ and
$0.05$ respectively.  This maximal attainable signal amplitude
decreases very strongly with the optical lattice depth, making the
signal in practise observable only up to a certain lattice depth which
depends on the interaction strength.

We have shown that the presence of a deep lattice has a dramatic
effect on the propagation of sound signals in the nonlinear regime:
Contrary to the uniform system, shock waves propagate slower than
sound waves, due to the negative curvature of the dispersion relation
in the lowest Bogoliubov band.  Moreover, nonlinearities can play a
role also at very small density variations and induce a saturation of
the sound signal, which goes along with dephased currents left beyind
the signal.  This effect has no analogue in the uniform case.  One
should also keep in mind, that the saturation effect is clearly
evident only if the shock waves move with a velocity much lower than
the sound velocity, which requires $\delta/U \ll 1/3$.  Obtaining a
clear saturation effect (small $\delta/U$) associated with a large
signal amplitude $\Delta n$ (large $\delta/U$) at fixed lattice depth
requires a compromise in the choice of lattice depth $s$ and
interaction strength $gn$.

This research is partially 
supported by the Mi\-ni\-ste\-ro dell'Istru\-zio\-ne,
dell'Uni\-ver\-si\-t\`a e del\-la Ri\-cer\-ca (MIUR), and by
the U.S. Department of
Energy through the Theoretical Division at the Los Alamos National
Laboratory.



\begin{references}
\bibitem{note2} The exact sound velocity can be extracted in this way,
thanks to our specific excitation method: a similar measurement
performed on a bright (grey) sound signal would lead to a slightly
higher (lower) value for the sound velocity.
\end{references}
\end{document}